# How to Compliment a Human - Designing Affective and Well-being Promoting Conversational Things

Ilhan Aslan[1,2], Dominik Neu[2], Daniela Neupert[2], Stefan Grafberger[3,2], Nico Weise[2], Pascal Pfeil[2], Maximilian Kuschewski[4,2]

[1] Huawei Technologies, Device Software Lab, Munich Research Center, Germany
[2] Augsburg University, Human-Centered Multimedia Lab, Germany
[3] AIRLab, University of Amsterdam, Netherlands
[4] Chair for Decentralized Information Systems and Data Management, Technical University Munich, Germany

aslan@hcm-lab.de, ilhan.aslan@huawei.com

**Abstract.** With today's technologies it seems easier than ever to augment everyday things with the ability to perceive their environment and to talk to users. Considering conversational user interfaces, tremendous progress has already been made in designing and evaluating task oriented conversational interfaces, such as voice assistants for ordering food, booking a flight etc. However, it is still very challenging to design smart things that can have with their users an informal conversation and emotional exchange, which requires the smart thing to master the usage of social everyday utterances, using irony and sarcasm, delivering good compliments, etc. In this paper, we focus on the experience design of compliments and the Complimenting Mirror design. The paper reports in detail on three phases of a human-centered design process including a Wizard of Oz study in the lab with 24 participants to explore and identify the effect of different compliment types on user experiences and a consequent field study with 105 users in an architecture museum with a fully functional installation of the Complimenting Mirror. In our analyses we argue why and how a "smart" mirror should compliment users and provide a theorization applicable for affective interaction design with things in more general. We focus on subjective user feedback including user concerns and prepositions of receiving compliments from a thing and on observations of real user behavior in the field i.e. transitions of bodily affective expressions comparing affective user states before, during, and after compliment delivery. Our research shows that compliment design matters significantly and using the right type of compliments in our final design in the field test, we succeed in achieving reactive expressions of positive emotions, "sincere" smiles and laughter, even from the seemingly sternest users.

# 1 Introduction

Imagine yourself getting ready to visit the latest exhibition in an architecture museum. You may plan to go there alone, with a partner, or with friends on your side. In any case you will spend some extra time, carefully choosing your clothes, putting on maybe one of your fashionable glasses, refreshing your hair color, trimming your beard, etc. until the image of yourself in the mirror pleases you and you believe that your appearance will fit and resonate with the specific occasion. Physical appearance is important for us humans as social beings who spend time in various self-styling and body shaping practices, including the consumption of beauty and fashion products, physical activities, healthcare and diets, and even surgery. Such practices are well described and studied in related literature and work on Somaesthetics [1]. We refer to interactive designs inspired by the interdisciplinary field of Somaesthetics, such as designs augmenting with technology all forms of self-experiencing and self-optimization practices simply as somaesthetic designs. Many such everyday practices relate to the social feelings of fitting into the surrounding environment and situation, feeling "integrated", and experiencing and producing resonance [2]. Feelings of resonance and its counterpart alienation are contemporary challenges in an accelerating social space with everything happening faster, being shorter lived, and opinionated with, for example, counts of likes (and dislikes) attached to statements and pictures one may share in social networks. Considering regular self-reflection and self-critique practices, mirrors are everyday objects affording such practices. Thus, many fellow researchers used and augmented mirrors for augmenting its functions and creating somasthetic designs.

For example, Dang et al. [3] have explored in related work an augmented smart mirror as part of a smart home, which is capable to recognize the user's emotions and reflects not only the physical appearance as traditional mirrors do but also the emotional expressions of the user by displaying ambient light in different colors; but, ambient and abstract information can be difficult to interpret for users. An alternative approach would be to provide explicit and concrete feedback, by using human language and direct speech. However, language is a very expressive modality and linguistic utterances and sentences need to be designed carefully (e.g. considering communication style and content) to produce feelings of resonance and affirmation and not alienation and mockery. For example, Perez-Marin and Pascual-Nieto [33] highlight that personality and mood conveyed by a conversational agent's style of communication influences the user's engagement with the agent measured in time spent interacting with the agent. Avula et al. [34] argue that timing of conversational interventions is essential to achieve user engagement with a conversational agent. Especially the design of first encounters with an agent [4] or unboxing experiences [5] can be a determining experience for the long-term perception of a design.

The motivation for the research in this paper was the assumption that a mirror would be an ideal smart thing to receive a compliment from and that receiving a compliment while using a mirror would be the ideal situation and time, since it is a moment of self-reflection for the user where they are in a (cognitive) process of "judging" themselves, a moment of exteroception and interoception in which one sees oneself seeing oneself. Using and catching this moment to affect a user's emotions in a positive manner was our main motivation.

However, giving a compliment is a complex "one-shot" problem. We needed to get the compliment design right the first time with the challenges that different users may prefer different types of compliments. To research (first) encounters with a complimenting mirror, we designed the "Complimenting Mirror" which gives compliments using speech synthesis (and displays the compliment also as text on the mirror). To understand compliment types and individual users' preferences, variations in preferences, opinions and reactions to a complimenting mirror, we first (i) conducted an online survey querying users opinions of receiving compliments and mirror practices and then (ii) conducted a Wizard of Oz (WoZ) study to observe how participants really react to a variety of different types of compliments presented by a smart mirror. The insights were used to (iii) implement a fully functional version of the Complimenting Mirror, which then was part of a month-long exhibition in an architecture museum. In the following we describe each step in detail. What we want to already share is that in contrast to the results we gained from the survey responses, in which only about 9% of the participants stated that they would enjoy getting a compliment from a mirror, in the field study we observed that the majority of users enjoyed the compliments from the mirror.

## 2   Background

Since the introduction of Affective Computing [6] many fellow researchers have studied how computers can utilize emotions and emotional mechanisms to improve human-computer interaction and allow interfaces to become more perceptive [7]. Researchers have studied, for example, how to enable affective loop experiences [8] and design interactions for an ``emotional living body" (e.g., [9,10,11]) by taking inspirations from Somaesthetics. Somaesthetics is a field that puts emphasis on the role of our human bodies in all aspects of our lives from the moment we start to move to the moment we become still, including the quality of experiencing and presenting ourselves to our surrounding world. Divers practices of self-styling, self-fashioning, or self-presenting are considered by Somaestehtics and categorized by how they may improve sensitivity and usage of our bodies as primary tools through which we interact and experience.

In contrast, the works of Calvo and Peters [12] on Positive Computing and Hartmut Rosa on a theory of resonance and alienation [2] focus on mental and social aspects, complementing the Somaesthetic perspective and lens to analyze human artifact interactions. Calvo and Peters [12] argue that research in Affective Computing and User Experience are the foundations for Positive Computing, which puts emphasis on technology's potentials to be directly designed for human well-being and enable human flourishing by applying user experience design methods and affective computing technologies to target "affective constructs", such as feelings of autonomy, competence, and relatedness. In the light of how IoT technologies and machine learning could change our everyday interactions with seemingly harmless objects, Resonance Theory, Positive Computing and Somaesthetics are valuable contemporary perspectives for designing interactions with technology, advocating sensitivity towards bodily, social and psychological consequences of technology usage and design. Our research is inspired and guided by all three, aiming to improve psychological well-being by designing for bodily self-appreciation and (positive social) resonance.

Fact is that individuals are rarely satisfied with how they look (e.g., [13, 14]). New factors introduced by social media, such as the mass distribution of filtered and altered self-images [15] negatively influence self-perception, as we are continuously exposed to unrealistic ideals. Consequently, there is an increasing dissatisfaction with the own body [16] with potentially unfiltered everyday experiences with mirrors causing negative affect and alienation rather than self-affirmation and resonance.

Experiences with interactive technology can indeed feel both alienating or resonating. For example, Sajadieh and Wolfe [17] critique and play with the idea of charming robots who apply seductive comments and compliments to evoke emotions in users and entertain bystanders. To create a resonating experience, culture and social conventions can serve as guiding constraints and provide friction for the design [18], and taking inspiration from interpersonal interactions in the design process is also recommended to prevent new automation designs to not be obnoxious [19].

Considering conversational user interface, the topic of task-oriented interfaces for both speech-based and text-based interaction is well studied with e.g. researchers exploring how mobile and multilingual conversational assistants can support tourists [37]. More recently Rapp et al. [35] have provided results from a systematic survey on the human side of conversational user interfaces, focussing on human-chatbot interaction. They describe for example why human's accept such technology and how they are emotionally involved. and highlight the complex ways in which emotions are involved in interactions with chatbots, the importance of theorization, well defined constructs and promoting human-centered design research.

For the topic of studying compliments, it is important to provide and use a definition of what a compliment is first:

*"A compliment is a speech act which explicitly or implicitly attributes credit to someone other than the speaker, usually the person addressed, for some "good" (possession, characteristic, skill, etc.) which is positively valued by the speaker and the hearer."* [20]

According to Pomerantz [21] the receiver of a compliment can accept or reject a compliment and they can agree or disagree with it. Furthermore, Pomerantz highlights the challenge of self-praise avoidance. We tend to avoid too positive statements about ourselves (for example by agreeing with a compliment) as these could be perceived as bragging. As a result a compliment receiver is under the pressure to react to the compliment properly.

To our knowledge, there is no research about the influence of compliments given by machines. However, there is a large body of related work about smiles and positive emotions. For example, Kleinke et al. [22] argue that eliciting smiles causes a positive feeling for humans. The smile must not be caused by a feeling, only the muscle movement is enough to evoke a positive emotion. Furthermore, seeing oneself smiling in a mirror will cause an affective loop enhancing the effect. This shows that trying to make people smile (especially in front of a mirror) is a simplified but reasonable approach when the goal is to improve their well-being. This general concept is implemented in the *HapinessCounter*'s [23] design, a smile encouraging design targeted mostly for users who live alone with limited social contact. Users are encouraged to smile to unlock the counter door and access important household items.

The project "*Hello Bot"* [24] is about a humorous looking robot that greets passersby and reacts to their smiles. The robot's goal is to improve the mood of its interaction partner. However, the robot does not adapt to the passersby, it interacts the same way with everyone. More recently Weber et al. [25] have presented a robot that learns the user's humor by telling different kinds of jokes and adapting the joke types based on recognizing users' laughter. In related research Ritschel et al. [26] have demonstrated how a robot that uses irony as a linguistic style element can be perceived as more likable. Smiles have also played a role in gamification approaches and related designs such as the *"SmileCatcher"* [27] game where the user has the task to make other people smile and capture it via a smart glasses device. Considering the in-cooperation of mirrors in interactive designs, related research exists for improving people's fitness [28], to discover health risks [29] and to support learners through augmentation [30].

Furthermore, we are not the first to work on the ideas of mirrors giving compliments. To launch a new brand Bimbo Brasil set up a mirror with a camera and speakers. In a car nearby a woman spoke compliments into a microphone which were then played by the speakers next to the mirror. It was their goal to compliment women and offer them for free a sample of their new product [31]. IKEA UK set up a mirror in one of their stores which complimented passing customers. The compliments were partly personalized, for example by mentioning a beard. It was IKEA UK's intention to boost the confidence of their customers as a lot of British are not satisfied with their appearance [32]. Both of these prototypes were created as a marketing campaign. Thus, in contrast to our work, the topic of designing an affective and well-being promoting conversational mirror was neither studied systematically and prototypes were not created through a human-centered design process, nor were any results or theorization to progress the field of interaction design research provided.

## 3 Phase 1: Human Mirror Usage and Complimenting Behavior

To inform our design decisions in the process of the creation and development of the Complimenting Mirror we started by distributing a survey using our own social networks, asking participants about their mirror usage, their perception of their appearance and their complimenting behavior. As it was not mandatory to answer each question the range of answers is 30 to 53. 67.17% of the participants were female, 32.86% male. 71.43% were under 29, 28.57% were 30 or older.

First, the participants answered questions concerning their mirror usage and their appearance. 43.47% of them stated that they are most likely to look into a mirror if they happen to see one (21 out of 39). 22.92% indicate that they sometimes feel insecure after looking into a mirror (11 out of 38). The participants were asked, which aspect of their appearance they liked most, the most common answers were as follows: eyes (46.67%), hair (16.67%), face (10%) and muscles(10%) (n = 30).

After that, questions about complimenting behavior followed. Most participants stated that they like receiving compliments (41.66%), with 27.08% stating to feel neutral about being complimented. However, the effect of compliments depends on the person giving the compliment. The participants indicate that they are most happy about compliments given by friends and family (70.84%), followed by compliments given by acquaintances (54.17%). Receiving compliments by strangers was stated as being least pleasant: only 41.66% state that they enjoy it.

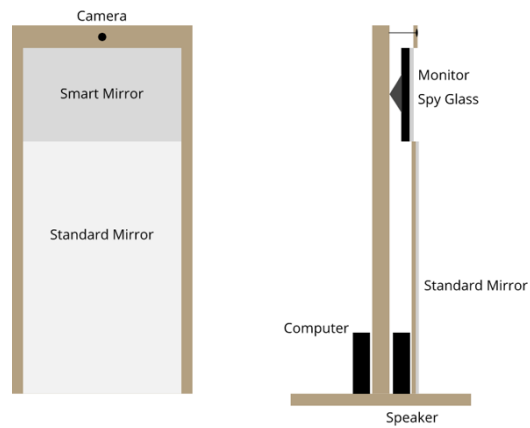

**Figure 1:** Construction of the prototype, the compliment is shown as text in the smart mirror section and the audio is emitted by the speakers.

We also asked the participant in which form they would like to receive compliments by a mirror. 34.29% said they would prefer compliments spoken in audio format and 42.86% would prefer written text. In order to respect both the visual and the auditive display options, we decided to integrate both output formats into our later prototypes. To answer the question of whether to use a female or a male voice for the complimenting mirror, we asked the participants which voice they prefer with voice assistants. 31.25% stated to prefer female voices, 9.38% prefer male voices and 46.88% do not prefer either of them. (12.5% chose the "other" option, n = 32). Consequently we decided to use a female voice for the text-to-speech output. We also asked the participant if they would enjoy receiving compliments given by a mirror. 54.29% indicated that they would not enjoy it, 28.57% were neutral and 8.57 % stated they would enjoy it.

Considering the design of the Complimenting Mirror we consequently decided to build a hardware prototype where users would be able to see a reflection of their whole body. The hardware setup is outlined in Figure 1 consisting of a monitor with spyglass, a regular mirror, a camera, a computer and speakers. Figure 2a shows a picture of the prototype in our "smart living lab".

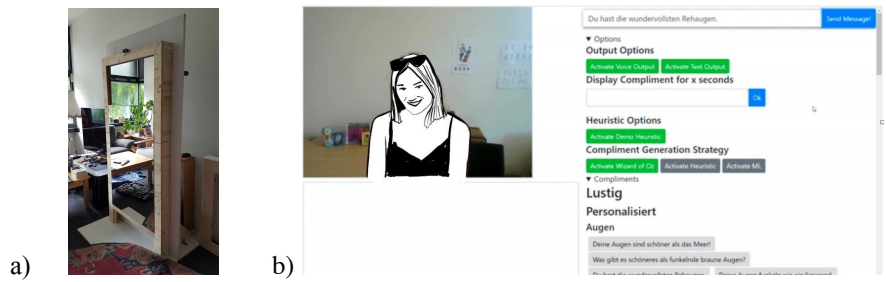

**Figure 2:** a) Photo of prototype of complimenting mirror used in the WoZ study. b) screenshot of experimenter's view on browser including live video of the participant (sketched for privacy preservation) and snapshot of the control interface.

## 4  Phase 2: Wizard of Oz Study

To observe real reactions to different kinds of compliments given by the Complimenting Mirror, we set up a WoZ experiment. Participants (n = 26, 8 female, 18 male, average age 21.77) were recruited around the university campus. To perform the experiment, we created a prototype application consisting of two Software components. The server component is a python application running on the smart mirror. The client component is a small web application used by the wizards to remote control the smart mirror. Experimenters can see the participants via its camera and send compliments either by directly typing them or by choosing one from a predefined list of choices (see Figure 2) . The server then outputs the chosen compliments in text and audio via the integrated screen and speaker. The client application records the participant's reactions and the wizard's interaction for further analysis. Compliments were pre-categorized as either personalized/specific or general/unspecific, and had the sub type earnest or humorous (see Table 1). The compliments were pre-categorized as following:

- Personalized/Specific: Compliments which refer to specifics of an individual's feature(s) and which can not be delivered to anybody
- General: Compliments which are not very specific and can be delivered to nearly anybody
- Earnest: Compliments using a factual language without any complementary metaphors.
- Humorous: Flirty compliments which aim to be amusing and use complementary metaphors as a style element

| *Personalized/Specific* | **PE** | **PH** |
|---|---|---|
| *General* | **GE** | **GH** |
|  | *Earnest* | *Humorous* |

**Table 1:** Possible permutations of compliment attributes.

The following are example compliments:

- PE: Your blonde hair is wonderful.
- GE: Wow, you look really good.
- PH: Holy cow, your glasses are looking great!
- GH: Your smile shines brighter than the sun.

### 4.2 Procedure

The hardware setup was distributed in two rooms. In one room the participant, the Complimenting Mirror, and the moderator were located. The experimenters (the "wizards") were located in a different room, remote controlling the Complimenting Mirror's behavior and observing the participant in front of the Complimenting Mirror.

Figure 3 presents an overview of steps in the study procedure. During the study each participant received 4 compliments (1xPE, 1PH, 1GE, 1GH) in counterbalanced order. After each compliment receiving experience we used questionnaire 1 to collect opinions from participants based on predefined questions and feedback using scores for disagreement-agreement in Likert scales for a set of 11 statements about the specific compliment and how participants felt about receiving the specific compliment. At the end of the study session, questionnaire 2 was applied and a semi-structured interview was conducted by the moderator, including the collection of opinions about what the participants liked/disliked about the Complimenting Mirror experience as a whole and what could be improved in future. To understand first impressions of users we performed a short interview with two questions after the participants received their first compliment from the mirror. We did this since first impressions can be lasting or defining the overall perception of a product's quality.

Both the interactions of the user with the mirror and the experimenters (the "wizards") input were recorded. During the study the experimenters had a set of predefined compliment choices at their disposal. But the experimenters could also use free text to produce specific compliments. Out of the 96 compliments delivered, 53 were different compliment sentences. Thus, we had a good variety of compliments. We also varied the use of general compliments. In the end, not one complimenting sentence was used more than 4 times. Compliments, where both presented as text in the mirror and through speech synthesis technology using a female voice. As mentioned before the output modalities and gender of the voice was chosen based on the survey results collected before the WoZ study.

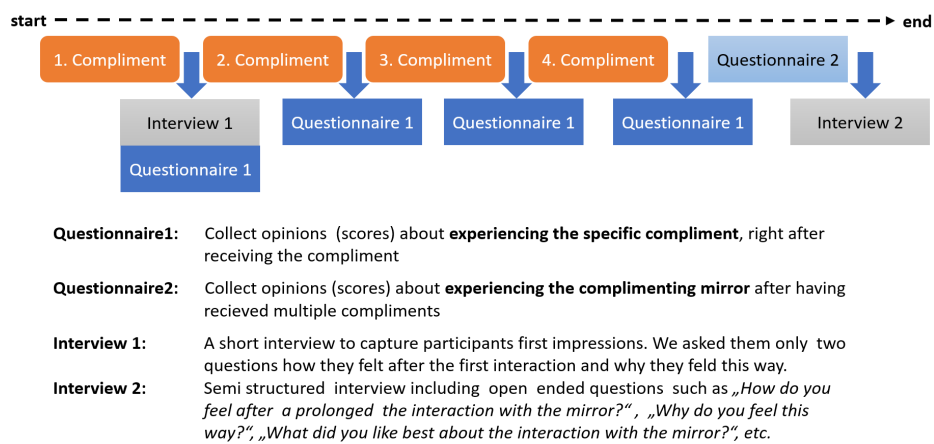

**Figure 3:** Overview of study procedure.

## 4.3 Observations

Most participants seemed open-minded toward the Complimenting Mirror and curious about the experiment. They seemed also nervous before they stepped in front of the mirror. After receiving compliments from the mirror, they seemed very excited about the experience they had just made and showed a great need to talk about it. There were also strong reactions to the specific compliments. Either they were pleasantly surprised at how well a compliment matched personal characteristics and thus wanted to know in detail how the mirror can detect such features. Or, conversely, they were upset when there was a compliment given that did not match the characteristics they expected. All participants assumed this version for the Complimenting Mirror to be a functional prototype. None of the participants suspected that the compliments were chosen or produced by humans in the background.

### 4.4 Results of the WoZ Study

**Results and discussion of interview 1:** The purpose of this first interview was to capture participants first impressions and initial mood. After receiving the first compliment, participants were asked how they felt after their first interaction with the mirror and why they felt that way. We analyzed participants' comments using a simple thematic analysis, identifying semantically related word counts. The main themes mentioned when participants described their feelings were, confusion, flattery, amusement, goodness, and interest.

For example, one participant who received the GH compliment *"your eyes are more beautiful than the sea"* stated to feel confused because the compliment was exaggerated and they were not sure if they could take it seriously. The comparison with the sea was perceived as strange. But on the other hand they said it was a compliment and made them smile even though they didn't believe the statement.

Another participant who received the PE compliment *"this is a mighty beard"* stated to never have received such a beautiful compliment and told us they felt great.

Another person who received the following PH compliment "You have wonderful deer-eyes" stated to feel good and amused. She stated the compliment was very personalized and detailed, and thus, not a standard compliment. Another participant who received the GE compliment "the haircut suits you" told us to feel amused, surprised, and pleased. They told us they didn't know what to expect and that they have never experienced something like this. They said thank you for the nice statement.

Some participants (n=8) mentioned being confused and somewhat caught by surprise. One participant provided the reason that it is weird when things start speaking with you and make compliments, others stated "because mirrors normally don't speak", "I wasn't sure what I was doing then I smiled and the mirror recognized and reacted to it", "It caught me by surprise and I didn't expect the statement", "Interesting that the mirror recognized my glasses [...] but since in theory a computer can not make subjective statements about aesthetics, the statement has not much value for me.", "I am surprised it recognized my hair color".

It seems that participants felt confused because they didn't expect the level of perception and social interaction from a thing. They seem to have questioned their own feelings (e.g., it felt human to some degree but it should not feel human because it was delivered by a computer, a thing).

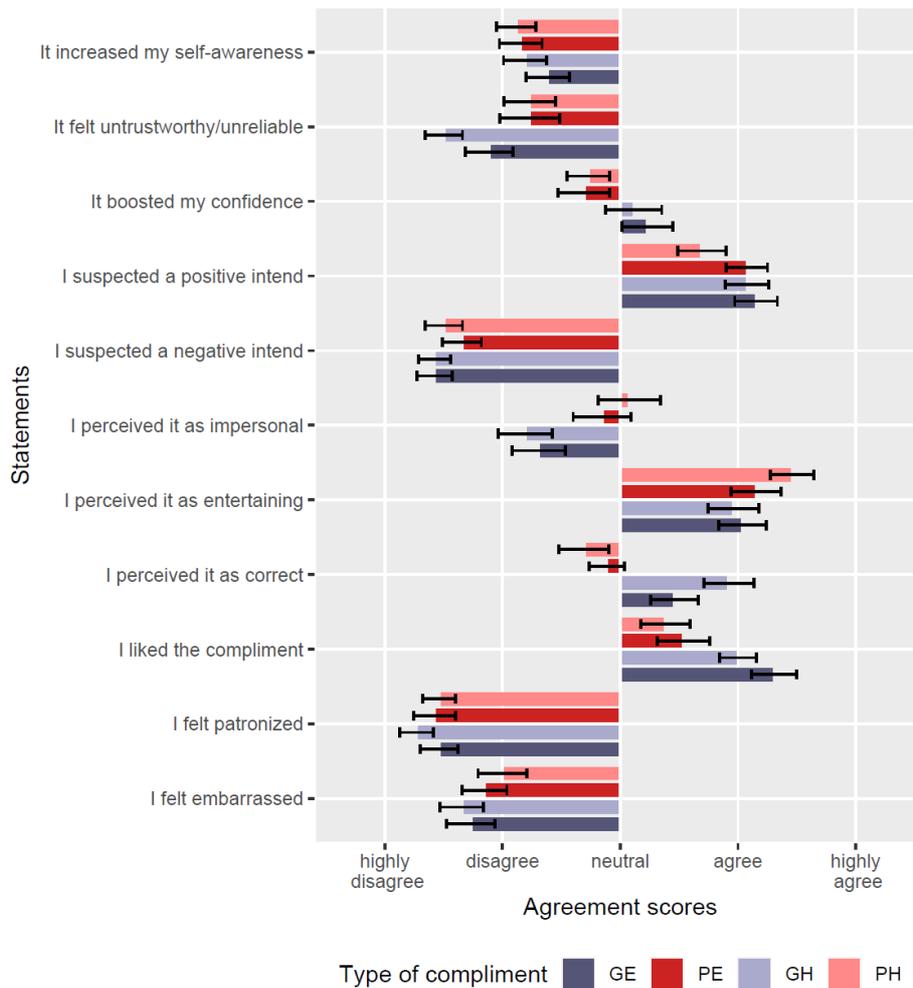

**Figure 4:** Overview of mean user agreements to statements about the received compliments. Error bars denote standard error.

**Results and discussion of questionnaire 1:** Figure 4 visualizes participants' self-reported agreement scores to the 11 statements we provided after each compliment receiving experience. The purpose of this questionnaire was to compare the effect of 4 types (GE, PE, GH, PH) of compliments. What we can see is that participants in general felt entertained by the compliments, they seem to have liked the compliment receiving experiences and they suspected positive intent behind the motivation of the compliments. GE compliments were liked most overall. Although PH compliments were perceived as most entertaining and PE compliments as most correct/accurate. The tendency of ratings is similar to what we would expect from receiving compliments from humans. Overall, most participants seem to have perceived the compliments as not embarrassing, not with negative intent or patronizing, even though in their understanding the compliments were given by the mirror, a thing/ machine. Thus, we can conclude that receiving compliments from a mirror are not undesired

Table 2 provides the significant results of the corresponding statistical analysis, considering the effect of compliment type (i.e., PE, PH, GE, GH) on the agreement scores. For six statements we observed significant differences, which are based on conducted repeated measurement ANOVA and post hoc tests. Type of compliment had a significant effect on participants' agreements considering (i) liking the compliment, (ii) correctness/suitability of the compliment, (iii) providing a confidence boost, (iv) being perceived as not personal, (v) assuming a positive intend, and (vi) perceiving a compliment as not earnest.

Thus, we conclude that the type of compliment matters. Our goal was to find out which type of compliments elicit the most positive reaction. The statistical analyses shed light to the results, beyond the comparison of mean values over all participants as it is mainly visualized in Figure 3. The post hoc tests tell us about significant differences in pairwise comparisons between types of compliments. For example, considering liking a compliment (first entry in Table 2), it matters significantly if the general or personalized compliment is humorous or earnest, and for earnest compliments it matters if it is personalized or general.

Furthermore, a compliment is perceived as more correct if it is earnest, independent of being general. The statistics also show that type of compliment matters in terms of perceived confidence, however there are no pairwise differences between the 4 types we compared. Looking at Figure 3 we can suspect the origin of the difference is between earnest and humorous, with earnest compliments boosting confidence more.

Considering Figure 4 compliments were overall not perceived as impersonal or too personal with ratings neither too low or too high. Since we used a direct speech style addressed directly to the participant, the participant will perceive a compliment as personal to some degree. While there are not significant differences between pairwise comparisons between the 4 types of compliments, there is a main effect of compliment type, we assume the significant effect of type of compliment on perceiving a compliment as impersonal is (with reference to Figure 4) very likely caused due to the difference between humorous and serious complements with humorous compliments (independent from the fact that they are general or personalized) being perceived as impersonal.

| Effect of compliment type on statements/opinions: | L Ratio and p-value | Significant post hoc tests |
|---|---|---|
| … liked the compliment | $\chi^2 = 18.1$, $p<.001$ | GH-GE: p<.001, PH-PE: p=0.3, PE-GE: p<.001, |
| … perceived as correct | $\chi^2 = 26.9$, $p<.001$ | PH-GE: p<.001, GH-PE: p<.001, PH-PE:p<001 |
| … boosted confidence | $\chi^2 = 7.9$, $p=.048$ | |
| … perceived as impersonal | $\chi^2 = 11.2$, $p<.001$ | |
| … positive intent | $\chi^2 = 16.2$, $p<.001$ | PH-GE: p<.001, PH-PE: p=.007, PH-GH: p=.007 |
| … untrustworthy/unreliable | $\chi^2 = 12.2$, $p<.006$ | GH-PE: p=.01, PH-PE: p=.009 |

**Table 2:** Summary of significant (i.e., p<.05) results of statistical tests, considering the effect of *compliment type* (i.e., personal-humorous (PH), general-humorous (GH), personal-earnest (PE), general-earnest (GE)) on participants' self-reports. See Figure 4 for corresponding mean ratings.

The type of compliment has a significant effect on the perception of positive intent with the significant differences between PH to any of the other 3 types. The combination of personal and humorous seems to have participants question the positive intent of the compliment compared to any of the other 3 types.

Last but not least, type of compliment has a significant effect on being perceived as not untrustworthy/unreliable. With pairwise significant differences between PE and GH, PE and PH. Thus PE (personal and earnest) compliments are perceived as least untrustworthy/unreliable. Consequently, PE (i.e., personal and earnest) compliments have achieved the most positive results and are suitable for a Complimenting Mirror to stimulate positive affective states.

**Results and discussion of questionnaire 2:** At the end of the lab study, we asked a set of additional questions to collect participants' more general opinions about the Complimenting Mirror. We wanted to know if they considered such a complimenting mirror acceptable in private and public spaces, if the complimenting experience was similar to perceiving compliments from a human or a computer, etc. Figure 5 provides the results in detail.

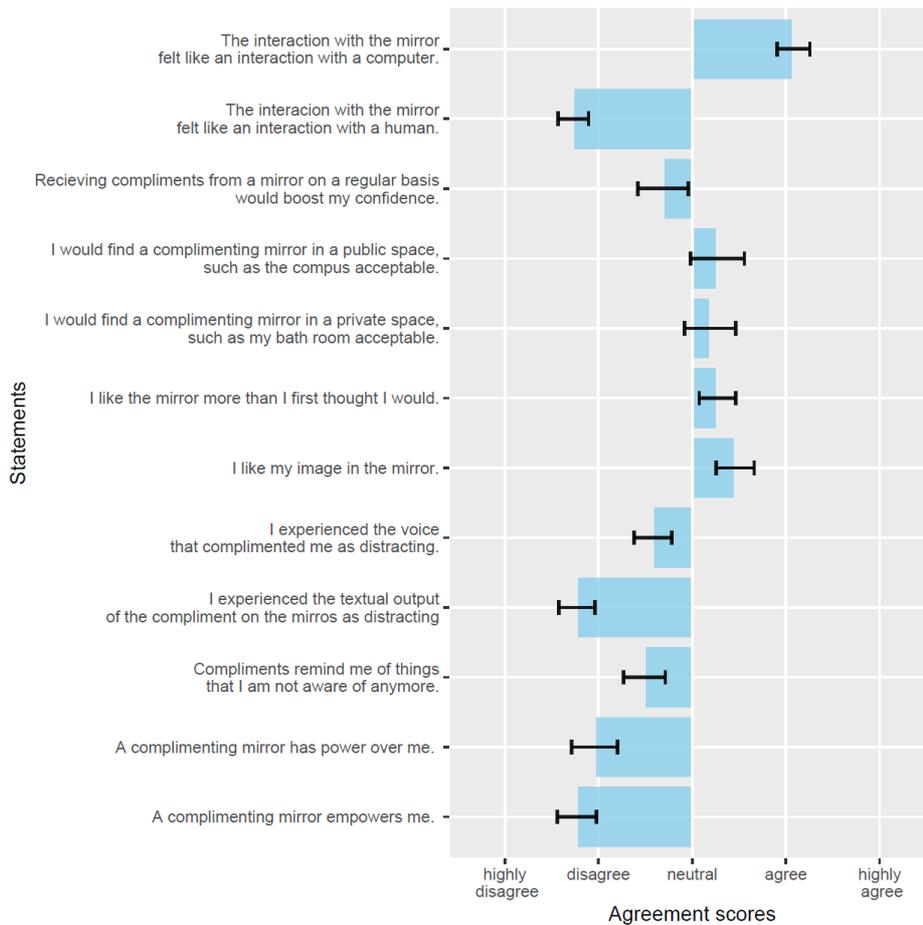

**Figure 5:** Overview of participants' opinions at the end of the study and after receiving different types of compliments from the complimenting mirror. Error bars denote standard error.

Participants agreed that the interaction with the mirror felt like the interaction with a computer and disagreed with the statement that it felt like the interaction with a human. Considering that this was a wizard of oz study with real humans choosing or providing the compliments, this result is somewhat surprising. We believe the opinions of participants could be different if a virtual avatar was displayed on the mirror delivering the compliments.

Participants' opinion on gaining a boost of confidence if they would receive compliments from a complimenting mirror was neutral to skeptical. We assume that when a mirror that provides a mix of different types of compliments will undermine a user's trust and the effect on confidence boost will be lost. Therefore, if the agenda is positive computing and improving users' well being the complimenting mirror should stick to giving personalized and earnest compliments.

All of our participants stated to be fine with their image in the mirror, we didn't have any extreme participants who openly stated to highly like or dislike their image in the mirror. We assume results could be different with such extreme cases. Participants reported no clear differences with regards to having a Complimenting Mirror in private space, such as the bath room in one's own apartment or having it located in a public space, such as a university campus. Both the textual and voice output of the Complimenting Mirror was not experienced as distracting, however participants seemed to be slightly more concerned with the voice output with some participants explicitly highlighting that the voice was not natural and human-like enough. A more human voice may also influence participants' opinion of the interaction feeling more like interacting with another human.

We also included questions regarding feelings of empowerment. Our goal is to design interactive artifacts which empower people. The two adjacent but somewhat contrary questions related to empowerment both received scores denoting disagreement of participants. Participants disagreed with both that a complimenting mirror has power over them or empowers them. We were hoping that compliments could be empowering by boosting confidence and providing approval.

**Results and discussion of the interview 2:** After the second questionnaire we started the last interview with the open question of *"How do you feel after a prolonged time of experiencing interactions with the complimenting mirror?"* followed by the question *"Why do you feel that way?"*. We had asked these two questions in the beginning of the first experience when participants had only experienced one type of compliment.

10 participants made statements related to improvement of their feelings, two participants stated to feel invigorated, other stated for example *"I feel good. The effect, the mirror had is larger than I spontaneously assumed."*, *"I feel a bit more self confident."*, *"I feel a bit better than before"*, *"I feel positive, surprised by the effect"*, *"I feel better than before, I feel a bit amused"*, etc.
The participants, who reported to have an improved feeling stated, for example the following to explain their reasoning: *"Compliments often have this effect"*, *"Compliments redirect my mind on positive things"*, *"Big effect of personalized compliments"*, *"It was fun to get compliments, especially from a mirror"*, *"Comments were relatively accurate"*, *"Because of the positive feedback"*.

Only one person provided this statement hinting at a deteriorated affective state *"I feel somewhat alienated"*, the same person had stated after the first round to feel confused and that it was weird to get compliments from things. They told us the reason for their feeling was to get a social reaction that made sense but at the same time knowing that an algorithm is producing the compliment. One person told us they feel confused because they don't see the beneficial utility of a complimenting mirror. Another stated to feel confused because the interaction is so unusual and that they don't like to talk with things. The rest of the participants told us to feel the same as before.

We also asked participants about what they liked and what could be improved. Nearly all participants stated that they like the fact that the mirror recognized details and that compliments were somewhat specific. Some (n=4) mentioned that they liked the fact that the interaction was automatic and that they didn't have to explicitly trigger it. Four participants mentioned the fact that the voice was not too robotic and 6 participants mentioned having also the text displayed to be able to read the compliment. One participant mentioned it was nice to see your own image and the compliment side to side to confirm the compliment was a good fit. One participant suggested that such a mirror would be good for someone with insecurities about their appearance.

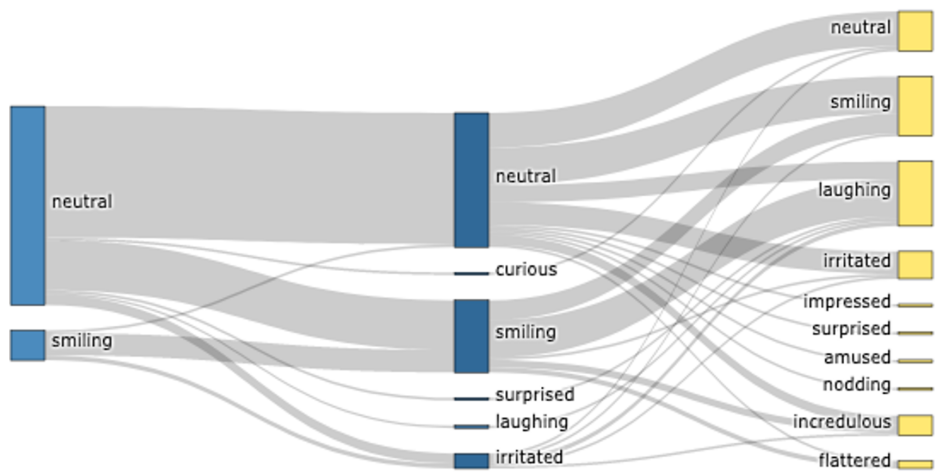

**Figure 6:** WoZ experiment: Emotions before, during and after receiving a compliment.

Considering improvements, it was mentioned that the voice still sounded a bit artificial, and that it would be nice if the mirror would greet someone first and was able to do small talk. Participants also suggested getting explanations such as visually highlighting what the compliments refer to. Furthermore, it was suggested to have the option to personalize the voice of the mirror, and to use emoticons in the text. One participant stated that one feels a violation of privacy, which needs to be addressed.

To achieve a more comprehensive analysis, we also analyzed the video recordings of the participants and labeled their emotional expressions before the compliment, during the compliment and after the compliment. Figure 6 depicts the change of affective state in detail. What we can definitely say is that the Complimenting Mirror stimulates positive affective expressions, such as smiling laughing which support many of the comments provided by the participants, especially the fact that 10 out of the 26 participants stated an explicit improvement of their affective states at the end compared to the first impression they had of the mirror.

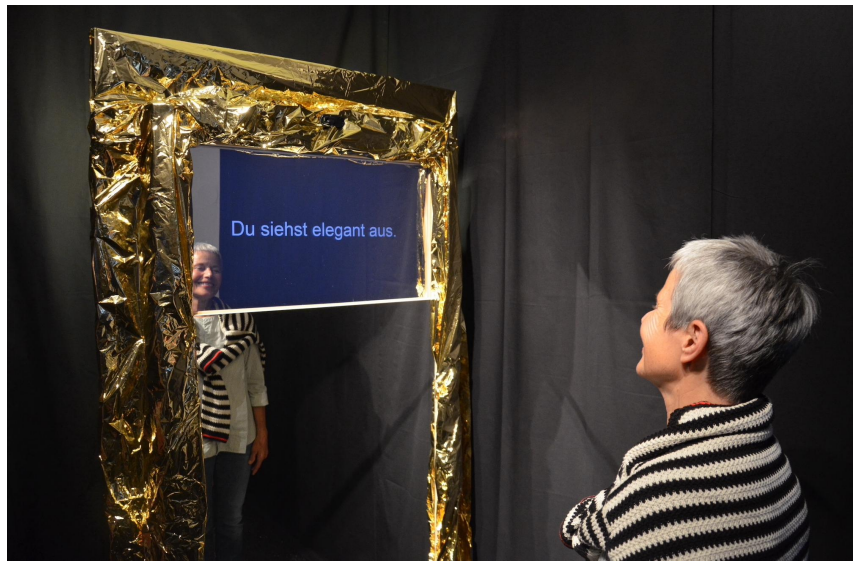

**Figure 7:** A visitor of the architecture museum receiving a compliment "You look elegant." during the field study.

## 5  Phase 3: Field Study at the Architecture Museum

In a final step of our research, we used the insights gained so far to finalize the functional version of the Complimenting Mirror, which was then exhibited without any administration in a Museum for a duration of a month. Figure 7 shows the mirror in the museum. The following changes were implemented in the fully functional design. As the personalized and earnest compliments were most appealing, we added a greater number of those to the compliment pool. Additionally, for some compliments we added distinctive formulations to get a wider range of different compliments, for example:

- You have beautiful glasses.
- You are wearing fancy glasses.
- Your glasses look good.

Though we did not rule out giving a differently formulated compliment about the same feature twice, the system avoids giving the exact same compliment twice. We did this, since in a museum visitor may stand longer in front of the mirror receiving multiple compliments. Furthermore, we excluded some of the humorous compliments and only kept the humorous ones which received a positive rating from the participants of the lab study

For the implementation we replaced the client software component that was used by the wizard in the wizard of setting with a module that uses the camera to automatically detect human faces and certain features. To accomplish this we used Microsoft Cognitive Services, which is a paid AI service. A frame from the video feed was sent to Microsoft Cognitive Services every 5 seconds. Recognizable features that these services deliver are hair color, gender, beard, glasses and smile. Based on the results from the Wizard-of-Oz experiment we implemented a rule-based heuristic that chooses fitting compliments for participants standing in front of the mirror. Our system also records videos of every interaction with users for post-hoc analysis to check if the fully functional version achieves similar positive results to the WoZ version in the lab study.

### 5.2 Field Study

We installed the Complimenting Mirror in Augsburg's Architecture museum as part of a new exhibition (see Figure 7). While the Mirror was exhibited for a month we were present only at the open day, since this was the day where most people would attend the exhibition, performing additional interviews with the visitors.

The mirror was placed in a darkened room and the participants only entered through one door. The opening was well attended, the news of a mirror that gives compliments spread quickly.

Before entering the mirror room, participants were asked for their consent in line with our institution's regulations. After using the mirror for as long as the participant wanted to, they were asked to fill out a short questionnaire. We asked them about their experience, their emotions, what they liked about the mirror and what could be improved in their eyes. Despite not every visitor who used the mirror being willing to postdoc fill out the questionnaire we collected a good number of questionnaires (n = 71). In addition, before, during and after a compliment was given, a short video was recorded to post-hoc analyze the facial expressions of the participants similar to how we did it in the lab study.

### 5.3 Field study results and discussion

The results of the analysis of the recorded video sequences (n=105) showed a consistent emotional expression change towards the positive, during the steps of our participants' interaction with the mirror (see Figure 8).

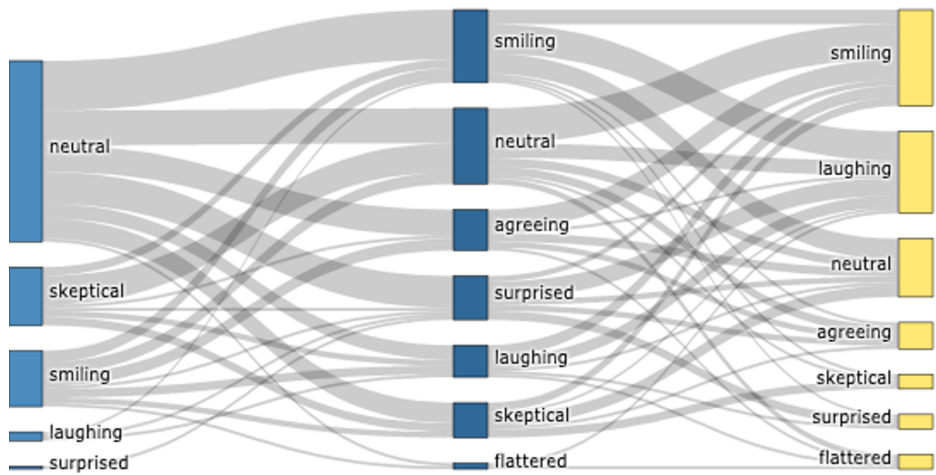

**Figures 8:** Field study: Emotions before, during and after receiving a compliment.

Already before going into detail, we can state that the participants enjoyed interacting with the mirror. During the field study, most participants received not only one compliment each. They stayed for an average of 2.7 compliments in front of the mirror. And that regardless of the fact that on the opening day in the museum many people queued up for the mirror and the pressure on the currently interacting visitor was high from time to time. This indicates that the participants generally perceive the interaction with the mirror as pleasant and engaging, contrary to the initial expectation (based on the online survey results asking for how people would feel getting compliments from a mirror), and did not feel uncomfortable.

Visitors of the museum, who used the mirror in the following days, remained on average for 6.8 compliments in front of the mirror. This means that without our supervision and without the sense of urgency caused by a queue of others waiting behind, people tend to interact with the mirror even longer.

As shown in Figure 9, initially most of the participants are neutral or rather skeptical. Only a few already face the mirror with a positive basic emotion. This is understandable, as the participants do not know what to expect and are exposed to a situation of potential social discomfort seeing a reflection of themselves. During the interaction, this initial emotion changes in almost all cases. After the compliment is given, most participants clearly show positive emotions, such as smiling or laughing.

The number of emotionally neutral participants decreases from 62 initially, over 26 during the interaction, to 20 after receiving the compliment. Similarly, the number of skeptical participants decreases from 20 at the beginning, over 12 throughout the interaction to 5 at the end. Complementary to this, the number of participants who smile or laugh towards the end of the interaction increases from 22 before, over 36 in between to more than half after (61 out of 105).

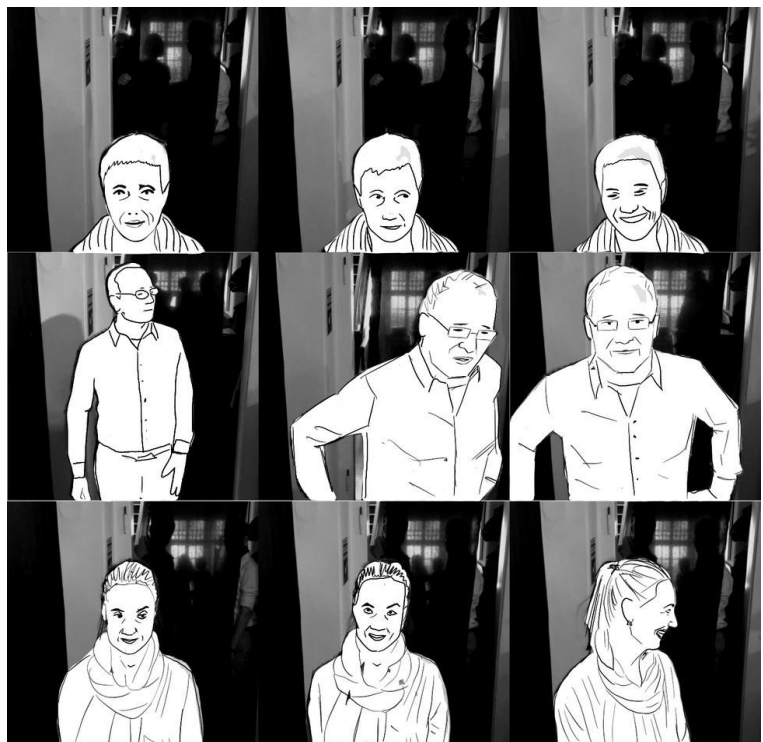

**Figure 9:** Field study: Pictures of the emotions before, during and after receiving a compliment.

This result that we draw from the video evaluation coincides with what the participants stated in the subsequent questionnaire: Nearly 80% of those who completed the questionnaire affirm that the interaction with the mirror was positive, for every fifth it stayed neutral and no one perceived the interaction as negative.

Considering most participants are not smiling or laughing initially, but start to do so in more than half of the cases, and remembering the muscle movement of these two facial expressions being key factors to improve humans' well-being, we conclude our design goal as confirmed.

In the questionnaire we asked the visitors how we could improve the mirror and one of the most frequent comments (5 out of 71) was considering the text-to-speech voice we used. It was perceived as unemotional and navigation-system-like. Furthermore, two participants stated that they would prefer a male voice. Four criticized that it takes too long until there is a compliment output. Two wondered if there could be not only compliments but negative comments as well which in their opinion would make the mirror more credible. Another suggestion by two participants was to make full conversations with the mirror possible. They wanted to be able to ask questions. Two other participants stated that they wanted to receive more personalized compliments.

In the field study, 62 of 105 people in the evaluated videos (59%) started their interaction with the mirror with a neutral expression and 22 people with a positive expression. After the interaction, 75 people (71%) showed positive reactions, 19% stayed neutral, and 4.7% showed signs of skepticism.

In comparison, in the WoZ lab study, 87% of the participants started the interaction in a neutral expression, and 13% with a positive expression. In this case, 60% ended the interaction with a positive expression, 26% kept a neutral expression, and 11% were even irritated or confused. Overall, we were able to achieve a positive change in emotion in both settings, though the measured positive impact was higher in the field study on average. The main differentiating factors were:

- The types of given compliments in the lab study were not optimized yet.
- The test participants in the lab study were observed in a lab setting. The test participants in the lab study did not socially interact with their peers. Instead, they were in a closed room together with a researcher.
- The test participants in the lab study were mostly students from the faculty of applied computer science, so the average age was lower, and the technical understanding of the inner workings of the mirror can be assumed to have been better than in the field study. Portela and Granell-Canut [35] have also argued that users with programming skills or a computer science background are inclined to be more skeptical when interacting with conversational agents (chatbots) in contrast to people with a humanities background.

While another difference was that some compliments in the lab study were ad-hoc defined by a human, and not generated by an algorithm. We assume that this fact did not contribute negatively to the emotional reaction shown by the participants, because a human is likely to be better at choosing a compliment with the intent to produce a positive emotional reaction than our algorithm.

Throughout the field study we observed some limitations that should be also mentioned. In the very beginning of the field study, the room with the mirror was accessible by just walking through an open door, groups of people gathered around it.

This led to confusion amongst the users because the compliment was generated not for the person who was closest to the mirror, but for the person with the strongest identifiable feature. As a result, the nearest participant assumed that the compliment referred to them and that their features were not detected correctly. We believe some form of highlighting (e.g. AR on the mirror or by linguistics) the addressee may be required in future work to prevent such misunderstandings. We immediately attached a curtain to the door and asked the participants to enter one by one. Another limitation of the prototype that we also observed is the necessity of standing frontally to the mirror. Some participants tested the mirror by showing only their profiles to it, some even were standing with their back to the mirror, expecting it to make a comment.

Due to the random selection algorithm we used to choose a feature to compliment on, it was possible that participants received multiple compliments pertaining to the same feature in a row. This effect was amplified because of the bad lighting conditions in the exhibition room, as there were less features our system could recognize with accuracy, and from which the selection algorithm could choose. This effect subverted the participant's expectation, as they mostly stood there to receive a compliment about another feature which can be seen in this questionnaire answer provided by one participant "Maybe a bit more variety, it was all about my hair".

This problem might be remedied by letting the system memorize which compliments about which features it gave to which users, and thus avoid giving similar compliments in a short amount of time.

A more important issue considering our design relates to its potential for discrimination or harm by putting unintentionally emphasis on vulnerable features, such as complimenting a balding man in public on his beautiful black hair, which socially is impolite and in most human-human situations conveys sarcasm. This has happened in the field study and if we had more older participants in the lab study we would have been able to spare the visitor from experiencing this potential humiliation.

## 6 Theorization

So far we have provided results and discussion of both design experiments, namely the WoZ lab study and the field study in the architecture museum. Our positive computing research and design activities are as mentioned before motivated by Affective Computing, Somaesthetics and Resonance theory. In the following we will use these perspectives with the aim to develop theorization for explaining user experience and design peculiarities (see Figure 10). A conventional mirror is an everyday artifact that is already used to conduct somatic practices to engage with the embodied emotional self. A mirror shows us how we appear to the rest of the world and what we look like when we are in a happy or bad mood, how our appearance changes over time when growing up and growing old, gaining weight, getting sick and healthy etc. An understanding of our self is consciously established through seeing and checking oneself (posing for oneself) over the course of our lives. How we see and feel ourselves is framed and contextualized socially (in a resonance environment) with e.g. the result that one feels like one is similar or has similarity (in potentially many ways) to the others (resonance) or that one is clearly different or has differences (alienation) to the others. As social beings there seems a need to check ourselves against the social environment, which motivates or even requests somatic practices to be performed on a regular basis. The UX of somasthetic designs are defined by the somatic practice, the self-feedback, and a resonance-alienation framework in which the self is continuously localized. Affective computing solutions are deployed in interactive designs which augment or disrupt somatic practices and thus can influence the embodied and emotional self of a person, and thus, how it localizes in the resonance-alienation framework. Figure 10 provides an actor network, explaining the relationships between the aforementioned perspectives (or abstract constructs). We believe actor network theory (ANT) is a suitable approach to describe the design specific dynamics.

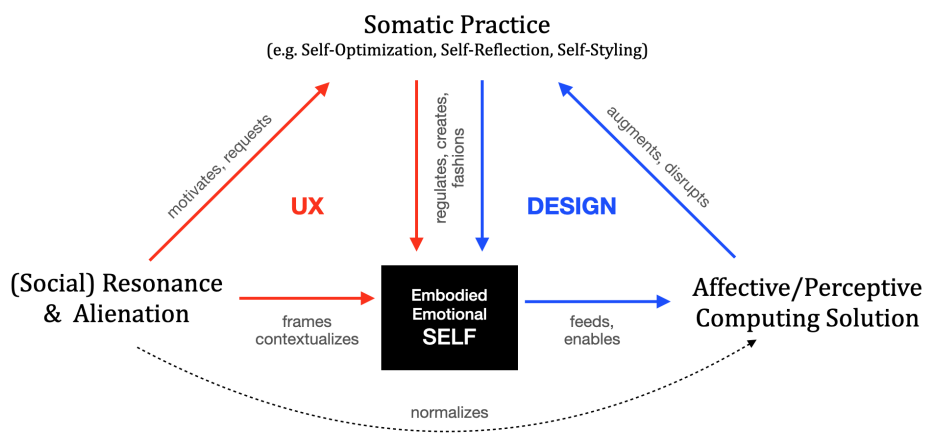

**Figure 10:** Actor network for theorization of the experience effects of interacting with the Complimenting Mirror and similar affective and perceptive solutions augmenting somatic practices, such as self-reflection and relation to self-experience/expression and Resonance/Alienation

## 7 Outlook

The progress in voice assistants is only the beginning of a journey and we as designers will be increasingly designing various forms of conversational user interfaces with things. We believe that conversational interfaces are powerfully affective and progress in technologies to recognize emotions in voice and language (e.g., [38]) and to express emotions in voice using emotional text to speech synthesis or voice conversion (e.g., [39, 40]) will drive the application and design studies of conversational and emotional user interfaces. Automatic recognition of affective behavior, such as laughter detection [25] and adaptation of content will become important to personal designs and mitigate some issues.

However, the truth is that it will probably get worse before it gets better, with emotional designs and adaptation techniques getting it too often wrong, conveying the wrong expression and content at the wrong time and context. Thus, more effort and care needs to be put in the design processes, such as adding mitigating rules in the inference algorithms, employing diverse design teams, and conducting pre-test with a broad spectrum of users to minimize any forms of unintended negative affect and affective damage to users. Having said this overall, we are happy that the compliments we designed and that were given by the Complimenting Mirror could achieve a positive emotional impact for so many users. The Complimenting Mirror was able to set a positive emotional context. Aspects that need more discussion is displaying an avatar or not, and using a more realistic and emotional voice or not, both these topics were brought up by participants of the lab study and at the museum. At this stage it is unclear if there would be more benefits if the design would be perceived as more human or not. For example, Deibel and Eanhoe [40] discuss in their book designing conversations with things, all the issues that can go wrong, including how voice and language communicates a persona and with that there can be potentially problematic associations of race, gender, class etc. For example, from which persona would people ideally want to get a compliment. The answer is not straightforward. The perceived persona matters and changes the context and impact of the compliment. The same issues arise when considering the use and design of an avatar with a visual presence. The more realistic and fine tuned the avatar appearance and voice will be, the clearer users will prescribe a persona etc. to the thing. I n our design we used a female sounding voice, and there is research indicating that females tend to give more compliments and receiving compliment from females is perceived as less problematic. However, we have had participants explicitly stating that they would have preferred a male voice. There is also research in nonbinary voice synthesis, which may also be appreciated by participants. Through the application of adaptive and implicit designs preferences could be identified, with the design trying out options and perceiving user reactions similar to work in human-robot interaction adaptations [25]. For the mirror experience, another option is that the voice could match the image of the user. There are voice conversion techniques being developed which would in future allow the mirror to use a voice very similar to the voice of the person standing in front of the mirror. Overall, this can be desirable since when you stand in front of the mirror you check your appearance and a voice complimenting your appearance that sounds at least a bit like your voice could be more effective in fostering self-appreciation and self-confidence. Other improvements could be achieved with progress in chatbot technology and chatbots being able to remember past interactions with individual users and not repeat themselves.

## 8 Conclusion

In this paper, we reported on a series of design experiments with the aim to enable a mirror to compliment visitors in a museum. Our experiments included a WoZ user study to explore the effects of various types of compliments and an implementation of a functional prototype, which was part of an exhibition at the Augsburg architecture museum for a month. We have described our design activities and observations in detail providing also a theorization. We believe the topic of designing well-being promoting and affective conversational things will become more and more important for designers with conversational and affective computing technology becoming increasingly accessible and powerful. We hope that fellow designers will benefit from our research to address the upcoming challenges and opportunities in design spaces where physical designs cross with conversational designs.

## 9 Acknowledgments

We thank all the volunteers who participated in our two studies and our online survey. Furthermore, we gratefully acknowledge the permission to exhibit our prototype in the Architekturmuseum Schwaben and thank the staff of the museum for their support. We want to also thank all the researchers of the Human-Centered AI lab in Augsburg who helped with the transport of the Complimenting mirror and special thanks to Dr. Tobias Baur, who provided and organized the very first prototype of the "digital mirror", on top of which the Complimenting mirror was built. The research described in this paper was conducted by all authors during their time as students or researchers at Augsburg University.